# Pulse-echo method can't measure wave attenuation accurately

Barnana Pal

Saha Institute of Nuclear Physics

1/AF, Bidhannagar, Kolkata 700064, India.

e-mail: barnana.pal@saha.ac.in

A number of techniques with different degrees of accuracies have been devised for the measurement of acoustic wave attenuation in solids and liquids. Still, a wide variation is observed in the attenuation values in different materials reported in the literature. Present numerical study based on a 'propagating wave' model analysis clearly shows that the attenuation constant obtained from exponential fitting of the echo heights in pulse-echo method differs from the exact value of intrinsic attenuation in the medium, even in the ideal situation of plane wave propagation without diffraction, dispersion or scattering.

Since the discovery of piezoelectric transducers, Pulse-echo method [1] has become the most popular experimental technique for the determination of acoustic velocity and attenuation in solids and liquids. Accurate measurement of attenuation becomes difficult due to reasons like diffraction, scattering and divergence of the ultrasonic beam. Efforts have been given to estimate relevant correction factors [1] and technical development as well [2-4]. Still, the common experience is that, though velocity values obtained are in good agreement with each other, attenuation values show wide variation and deviates much from theoretical calculations. To be specific, we consider the data available in the literature for pure water [2,3,5-8]. These are presented in Table-1. A recent work by R Martinez et. al. [7] reports a comparison between the through-transmission and pulse-echo techniques for measuring attenuation in tri-distilled water at $25^{o}C$ and $19.8^{o}C$. They observed high variation in the value of α measured at short distances from the transducer whereas at long distances measurements in the two methods produce comparable data but, the value differs widely from other reported values. Secondly, their result shows higher wave amplitude for pulse-echo method at far zone and attenuation value obtained from through transmission method is found to be a little higher than that obtained by pulse-echo method. P K Dubey et. al. [3] developed a high resolution technique based on very accurate measurement of echo amplitudes in pulse-echo set up and obtained a value for attenuation in distilled water at $34^{o}C$ comparable to the result reported by J M M Pinkerton [5]. In this communication we present a 'propagating-wave' model approach [9] to analyze the pulse-echo response and from numerical calculation we show that the attenuation constant ($\alpha_m$) obtained from an exponential fitting of the observed echo amplitudes deviates from the exponential decay constant (α) of the freely propagating wave.

Table I : Literature values of attenuation constant in pure water.

| Data Source | $\alpha/f^2 \times 10^{15}$ (np sec$^2$/cm) | Frequency Range (MHz) | Temp. ($^0$C) |
|---|---|---|---|
| J M M Pinkerton, 1947 [5] | 0.22 | 7.37-66.10 | 25 |
| E G Richardson, 1962 [6] | 1.0<br>0.4 | 1.0<br>10.0 | 20 |
| X M Tang et.al., 1988 [2] | < 1.0 | 0.2-1.5 | 20 |
| P K Dubey et.al., 2008 [3] | 0.21 | 10.0 | 34 |
| R Martinez et.al., 2010 [7] | 44.17 | 1.0 | 25 |
| B Pal & S Kundu, 2013 [8] | 75.42<br>16.93 | 1.0<br>2.0 | 25 |

Bolef and Miller introduced the 'propagating wave' model [9] to explain the response of an ultrasonic resonator consisting of the transducer bonded to the sample. The model has been further modified [10,11] to explain the responses observed in pulse-echo, continuous wave (cw) and others combining the two, considering plane wave propagation along the length of the sample. Here we will present a brief description of the model for clarity. The particle displacement $u$ at any position $x$ measured along the length inside the medium at time $t$, during propagation of a monochromatic wave of frequency $\omega=2\pi f$ with a velocity $\upsilon$ is represented as,

$$u(x,t) = a\, e^{-\alpha x} e^{i(\omega t - kx)}, \qquad (1)$$

$\alpha$ being the attenuation co-efficient and $k$, the propagation constant. The wave amplitude decays exponentially from the initial value '$a$' at source point with distance during free propagation. In a medium of finite length $l$, the wave suffers successive multiple reflections at the end faces aligned perfectly parallel to each other and perpendicular to the direction of propagation. Under the assumption of linear superposition, the steady state particle displacement can be written as [11],

$$u_m(x,t) = A e^{i\omega t} \qquad (2)$$

where the modified wave amplitude $A$ is given by,

$$A = \frac{a[e^{-ik'x} + re^{-ik'(2l-x)}]}{[1 - r^2 e^{-ik'l}]}, \qquad (3)$$

where $k' = k - i\alpha$ and $r$ is the reflection coefficient at the boundary. Considering the real part of $u_m(x,t)$, an alternate expression of $A$ may be written as,

$$A = \frac{ae^{-\alpha x}[1 + r^2 e^{-4\alpha(l-x)} + 2re^{-2\alpha(l-x)}\cos 2kx]}{[1 - r^2 e^{-2\alpha l}]}, \tag{4}$$

The above expression shows that the amplitude at any position $x$ inside the medium depends on $r$, $k$, $\alpha$, and $l$ in a complicated manner. Further, in pulse-echo experiments, short-duration carrier-frequency rf pulses are used to probe the system. Such pulses may be considered as the linear superposition of a large number of continuous waves propagating in the medium with frequencies and amplitudes determined by the Fourier spectrum of the pulse signal [12]. Thus, to get the response under pulse excitation, we have to sum up over all the relevant frequency components with appropriate phase velocities contained in the bandwidth of the transducer loaded with the sample and the associated electrical circuit. Thus, resultant particle displacement in a bounded medium under pulse excitation will be,

$$U(x,t) = \sum_\omega u_m(x,t) \tag{5}$$

In case of wave propagation in a medium of finite length $l$ bonded to the transducer at $x = 0$, the response will be proportional to $U(0,t)$. We also assume that, all of the frequency components contained in the transducer bandwidth have same initial amplitude and they propagate with the same phase velocity. For numerical evaluation of the exciting pulse and pulse-echo responses, we choose the following parameters: carrier frequency $f_c = 10$ MHz, and frequency bandwidth 9.5-10.5 MHz, width of the pulse 2 $\mu$s, pulse repetition time 50 $\mu$s, intrinsic attenuation constant $\alpha = 0.1$ $\mu s^{-1}$ and propagation velocity c = $10^6$. Responses have been evaluated for different values of $r$ ( $r$ = 0.6, 0.7, 0.8, 0.9, 1.0) and $l$ ( $l$ = 2.0, 2.25, 2.5, 2.75, 3.0. 3.25, 3.5, 3.75, 4.0). In each case attenuation constant $\alpha_m$ is determined from the exponential fitting of the evaluated echo heights. The exciting pulse and typical echo patterns in case of sample length $l$ =2.5 cm for r = 0.6, 0.8 and 1.0 are presented in fig 1(a), (b), (c) and (d) respectively. In fig 1(d) the height of the first echo is higher than the input signal, which is never observed in real experiments because of conversion loss of the transducer at the transmitting boundary. The values of $\alpha_m$ obtained are indicated in the figure and it is seen that even in the ideal case, where there is no diffraction or divergence of the propagating wave, the measured attenuation $\alpha_m$ differs significantly from the intrinsic attenuation constant $\alpha$ taken to be 0.1 $\mu s^{-1}$ for numerical evaluation.

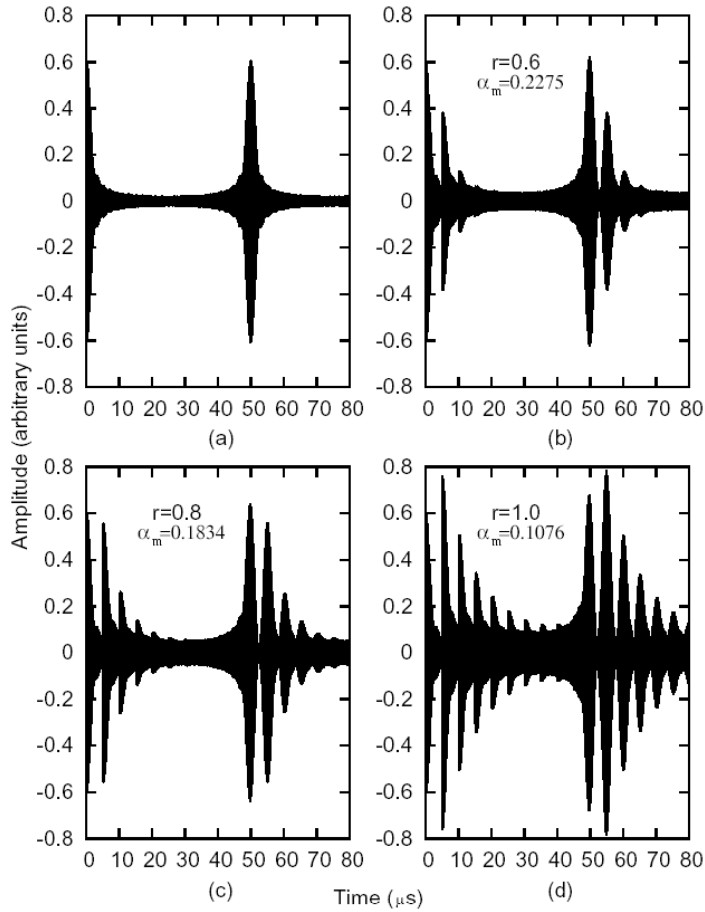

Figure1: Exciting input signal (a) and numerically calculated pulse-echo responses for (b) r=0.6, (c) r=0.8 and (d) r=1.0. Sample length l is taken to be 2.5 cm.

Figure 2 shows the variation of $α_m$ with $r$ for $l = 2.5$. $α_m$ is found to be significantly higher than α for lower values of $r$ but the difference is reduced with the increase of reflectivity at the end faces. Still, even for perfectly reflecting faces, i. e. for $r =1.0$, $α_m$ is higher than α.

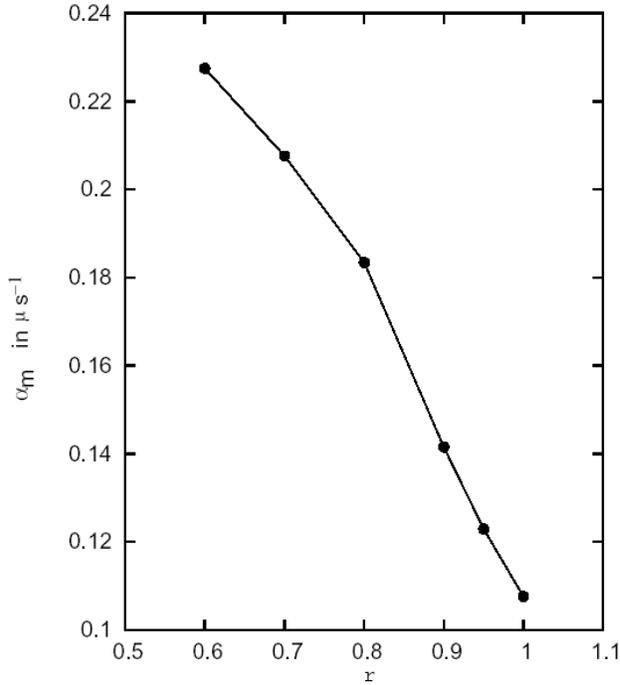

Figure2: Measured attenuation $\alpha_m$ in $\mu s^{-1}$ vs. reflection coefficient r for l=2.5.

In figure 3 we present the variation of $\alpha_m$ with change in the sample length $l$ for $r = 0.8$ and 1.0. We see that, for small sample lengths, $\alpha_m$ shows random variation with $l$ and $\alpha_m$ may be less than $\alpha$ for large r ($\approx$ 1.0). With the increase of $l$, $\alpha_m$ approaches a steady value and again, the steady value is higher than $\alpha$ even for $r = 1.0$. In actual experiments, one can increase $l$, but up to a certain extent depending on $\alpha$, so that a reasonable number of echoes are detectable. Moreover, since $r$ is not known, one cannot be sure how close is the value of $\alpha_m$ to $\alpha$.

Equation (4) shows that the amplitude of the Fourier component waves of the input pulse is modified due to multiple reflection at the end faces and the steady state wave amplitude at any position $x$ inside the medium at time $t$ depends on $r, k, \alpha$, and $l$ in a complicated manner. So, it is expected that measurement of attenuation from echo heights in pulse-echo method will depend on these factors. The present numerical results agree with this interpretation. Also, in real experiments, it is reported in the literature that measurement of attenuation constant depends very much on experimental condition [2,5,7]. The observation of Martinez et. al. [7] is important in this context. They obtained huge difference in the attenuation value measured by through-transmission and pulse-echo method at small distances from the transmitting transducer, whereas at far distances the values are comparable. Attenuation obtained by pulse-echo method is a little lower than that obtained from through-transmission method, which is not in consistent with our present study, but this may arise due to errors in the measurement.

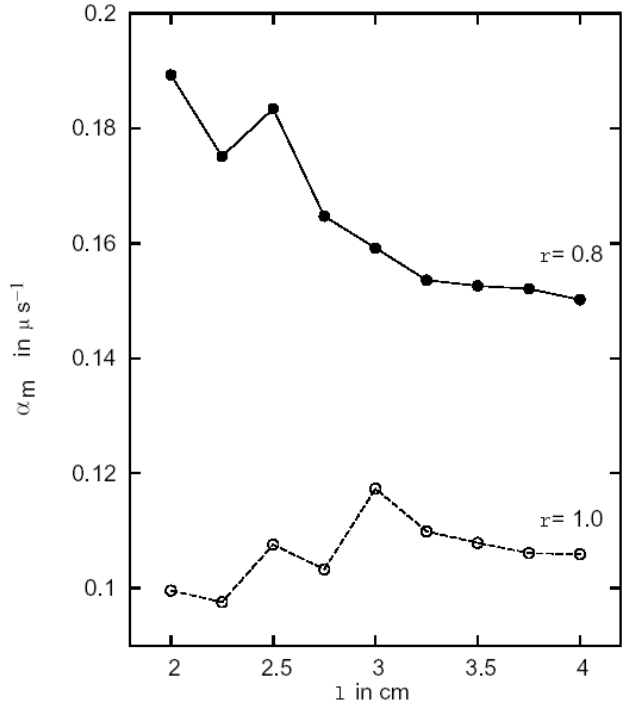

Figure3: Measured attenuation $\alpha_m$ in $\mu s^{-1}$ vs. length $l$ in cm for $r= 0.8$ and $1.0$.

Thus, the present study shows explicitly that, even in the ideal case of plane-wave propagation where there is no dispersion, diffraction or scattering loss, attenuation measurement in pulse-echo set-up differs significantly from the intrinsic wave attenuation in the medium. The method is useful only to have an estimate for the upper limit of the desired attenuation constant. Accurate determination of α is possible from the pulse-echo signal with the application of a Fourier transform computational method as has been adopted for the study of the dispersion of ion-acoustic waves in plasma [11]. This requires Fourier analysis of the echo-pattern obtained for several known sample lengths under similar experimental conditions so as to keep *r, k* and α constant. Knowing the amplitudes obtained for various sample lengths for a particular component wave a computational method of parameter fitting using relation (4) may be adopted to find out accurate values for *r, k* and α.

--------------------------